\documentclass[conference]{IEEEtran}

\usepackage[utf8]{inputenc}
\usepackage[T1]{fontenc}
\usepackage{cite}
\usepackage{amsmath,amssymb,amsfonts}
\usepackage{graphicx}
\usepackage{booktabs}
\usepackage{array}
\usepackage{multirow}
\usepackage{url}
\usepackage{xcolor}
\usepackage{tikz}
\usetikzlibrary{arrows.meta,positioning,fit,calc,shapes.geometric}
\usepackage{listings}
\usepackage{algorithm}
\usepackage{algpseudocode}
\usepackage{enumitem}
\usepackage{balance}

\definecolor{cbenign}{HTML}{0072B2}
\definecolor{cinj}{HTML}{E69F00}
\definecolor{chij}{HTML}{D55E00}
\definecolor{cfail}{HTML}{009E73}
\definecolor{codebg}{HTML}{F6F6F6}

\lstdefinestyle{steptext}{
  basicstyle=\ttfamily\scriptsize,
  columns=fullflexible, keepspaces=true, breaklines=true,
  breakatwhitespace=false, showstringspaces=false,
  frame=single, framerule=0.3pt, backgroundcolor=\color{codebg}
}

\newcommand{\yes}{$\checkmark$}
\newcommand{\no}{$\times$}
\newcommand{\partialmark}{$\circ$}
\newcommand{\lbl}[1]{\texttt{#1}}
\newcommand{\driftnet}{DriftNet}

\begin{document}

\title{DriftNet: A Dual-Head Trajectory Transformer for Detecting and Localizing Prompt Injection in LLM Agents}

\author{\IEEEauthorblockN{Asif Pinjari}
\IEEEauthorblockA{\textit{School of Informatics, Computing,}\\
\textit{and Cyber Systems}\\
Northern Arizona University\\
Flagstaff, AZ, USA\\
ap3929@nau.edu}
\and
\IEEEauthorblockN{Mithun Paul Saint-Germain}
\IEEEauthorblockA{\textit{School of Informatics, Computing,}\\
\textit{and Cyber Systems}\\
Northern Arizona University\\
Flagstaff, AZ, USA\\
mithun.paul@nau.edu}}

\maketitle

\begin{abstract}
When an indirect prompt injection succeeds against an LLM agent, the compromise is visible in the agent's own behavior: a benign prefix of tool calls, a poisoned observation, and a suffix of actions that serve the attacker. An operator needs three facts: where the attack entered, which steps it corrupted, and whether apparent poison was resisted. Existing systems return either a whole-trace verdict or a single unsafe index. We present DriftNet, a dual-head trajectory Transformer that reads a logged tool-call trajectory and answers all three questions in one forward pass: one head classifies the trajectory as compromised or not, and a second assigns every step one of four labels (benign, injection point, hijacked, failed injection). To our knowledge it is the first supervised detector to produce this joint output. A frozen sentence encoder and four identity-free world features embed each step; the trained trunk, under two million parameters and optimized with a class-weighted joint objective over both heads, needs no access to the agent's model. On the task-disjoint split of the AgentDrift benchmark (12,536 trajectories, 71,024 labeled steps), with a 20-configuration sweep bounding hyperparameter sensitivity to 0.011 F1 and the test part evaluated exactly once, DriftNet reaches trajectory-level F1 of 0.983, exact injection-point recovery on 98.7\% of attacked trajectories, hijacked-span IoU of 0.979, zero flags on 218 resisted attacks, and 2.9\% flags on hard negatives. A surface baseline retrained on the identical split recovers 11.1\% of partial hijacks and 17.1\% of delayed executions; DriftNet reaches 98.6\% and 93.2\% while lowering every false-alarm rate. Reading all 26 residual errors shows that most misses trace to trajectories whose labeled injection observation carries no legible instruction, and we report the benchmark's measured world-identity regularity alongside the results.
\end{abstract}

\begin{IEEEkeywords}
LLM agents, prompt injection, trajectory analysis, sequence labeling, anomaly detection, agent security
\end{IEEEkeywords}

\section{Introduction}
\label{sec:intro}
LLM agents act on the world by interleaving reasoning with tool calls. Following the pattern popularized by ReAct~\cite{react2023}, every step couples a reasoning thought with a tool invocation, its arguments, and the returned observation, and such agents already read inboxes, move money, browse the web, edit code, and retrieve patient records. Every observation the agent reads is also an attack surface. Indirect prompt injection~\cite{greshake2023} hides instructions in content the agent will later retrieve; since a language model draws no hard line between data and instructions, the agent can adopt the planted goal as its own. The threat is measured, not hypothetical: a ReAct-prompted GPT-4 follows injected instructions in roughly a quarter of InjecAgent's test cases~\cite{injecagent2024}, Agent Security Bench reports attack success above 80\% for some backbones~\cite{asb2025}, and stage-level tracking shows injected payloads surviving across surfaces and agent boundaries in multi-agent pipelines~\cite{killchain2026}.

When an injection succeeds, the compromise has a shape. Execution begins benignly, a poisoned observation arrives mid-task, and the steps that follow drift away from the user's goal toward the attacker's. Detecting this is a sequence problem: each step must be judged against the task, the world the agent operates in, and everything that came before. Localizing it is what makes detection actionable. An operator confronted with a flagged trajectory needs to know the step where the attack entered, the span of actions it corrupted, and whether apparent poison was actually resisted, because those three facts decide what to roll back, what to audit, and which content source to distrust.

Existing systems do not produce this output. Live-loop defenses veto or re-derive single actions and assume control of the running agent~\cite{zhu2025melon,an2025ipiguard,attriguard2026,dreamguard2026}; internal-state probes and attention localizers require white-box access to the agent's model~\cite{hiddenstate2026,basis2026,attnlocate2026}. Detectors that do read completed trajectories emit either a single verdict on the whole trace, with resisted injections folded into the safe class~\cite{yuan2024rjudge,li2026atbench,liu2026agentdog}, or a single index: the first unsafe action~\cite{zheng2026stepguard}, the one mutated step~\cite{chen2026tracesafe}, the divergence onset~\cite{felicia2026stepshield}, the error step~\cite{trajad2026}, or the root-cause component~\cite{chen2026attribution}. A single index cannot represent an agent that complied for two steps and recovered, an execution delayed past benign steps, or an injection that was seen and refused. What has been missing is a detector whose output is the full picture: a trajectory verdict together with an injection-specific label on every step.

This paper presents \driftnet{}, a detector with exactly that output, trained and evaluated on the AgentDrift benchmark~\cite{agentdrift2026dataset}, whose 71{,}024 step labels make dense, injection-specific supervision of this task available for the first time. \driftnet{} is deliberately small. Each step is serialized to text and embedded by a frozen sentence encoder; four world-grounded, identity-free features recover what semantics cannot see, namely whether the step's action targets a recipient outside the user's known world; and a Transformer encoder of at most three layers reads the sequence with two heads, one pooling into a trajectory verdict, one labeling every step as \lbl{benign}, \lbl{injection\_point}, \lbl{hijacked}, or \lbl{failed\_injection}. The trained trunk is under two million parameters, three orders of magnitude below contemporary LLM-scale guards~\cite{zheng2026stepguard}, and it consumes only the logged trajectory: no agent internals, no re-execution, no live loop.

The evaluation is designed to be hard to fool and easy to trust. All experiments use the corpus's task-disjoint split, in which no task template is shared between training and test, so template memorization cannot inflate the numbers. A 20-configuration random sweep establishes that the result does not depend on a lucky configuration (every draw lands within a 0.011 band of validation F1). The held-out test part is evaluated exactly once. And the surface baseline we compare against is retrained on the identical split, making the comparison like for like. Under this protocol \driftnet{} reaches trajectory-level F1 of 0.983; it recovers the exact injection-point set in 98.7\% of attacked trajectories and the hijacked span at mean IoU 0.979; it flags zero of 218 resisted attacks and 2.9\% of hard negatives. Where the surface baseline collapses on the benchmark's two stealthy compliance patterns (11.1\% recall on partial hijacks, 17.1\% on delayed executions), \driftnet{} reaches 98.6\% and 93.2\%. Reading all 26 residual errors individually shows that they are confident rather than marginal, that the misses concentrate on delayed executions, and that in nine of twelve misses the labeled injection observation contains no legible instruction at all, pointing to residual generation noise in the corpus rather than to evidence the detector failed to read.

We also state plainly what these numbers do not establish. The corpus is synthetic and single-generator, and the dataset paper measures a world-identity regularity in it that a lookup exploits to 86\% binary accuracy; our world features are immune by construction, but the text embeddings are not, and Section~\ref{sec:limitations} discusses what that does and does not explain. The claim of this paper is scoped accordingly: on the first benchmark dense enough to supervise the task, joint trajectory classification and four-way step labeling is learnable to high fidelity by a small, cheap, model-agnostic detector.

Our contributions are as follows.
\begin{itemize}[leftmargin=*, itemsep=1pt]
\item \textbf{\driftnet{}.} A dual-head trajectory Transformer that, to our knowledge, is the first supervised detector to jointly classify a tool-call trajectory and label every step four ways, locating the injection point, the hijacked span, and resisted injections in one forward pass from the log alone (Section~\ref{sec:method}).
\item \textbf{Strict localization metrics.} Injection-point exact-set match and hijacked-span IoU, which expose localization failures that per-class F1 hides (Section~\ref{sec:metrics}).
\item \textbf{A trust-preserving protocol.} Task-disjoint training, a robustness sweep, a single test evaluation, and a like-for-like retrained baseline (Section~\ref{sec:protocol}).
\item \textbf{Results and analysis.} High-fidelity detection and localization on AgentDrift with pattern, domain, and family breakdowns; an exhaustive error analysis with a calibration caution; and an honest account of the benchmark's measured artifacts (Sections~\ref{sec:results} to~\ref{sec:limitations}).
\end{itemize}

The remainder of the paper proceeds from related work (Section~\ref{sec:related}) through formulation (Section~\ref{sec:problem}), the benchmark and split (Section~\ref{sec:benchmark}), the architecture (Section~\ref{sec:method}), and the protocol (Section~\ref{sec:protocol}), to results (Section~\ref{sec:results}), error analysis (Section~\ref{sec:errors}), discussion (Section~\ref{sec:discussion}), limitations (Section~\ref{sec:limitations}), and future work (Section~\ref{sec:future}).

\section{Related Work}
\label{sec:related}

\subsection{Indirect Prompt Injection and Agent Attack Benchmarks}

Prompt injection was first characterized as an attack class on instruction-following models~\cite{perez2022ignore} and then shown to compromise deployed LLM-integrated applications through content the model retrieves rather than through the user's prompt~\cite{greshake2023}. Liu et al.\ formalized the attack family and benchmarked defenses at the prompt level~\cite{liu2024formalizing}. For tool-using agents the threat is measured by live attack benchmarks: InjecAgent reports that a ReAct-prompted GPT-4 follows injected instructions in roughly a quarter of cases~\cite{injecagent2024}, AgentDojo executes attacks inside a stateful tool environment~\cite{agentdojo2024}, Agent Security Bench reports attack success above 80\% for some backbones~\cite{asb2025}, and AgentDyn moves the evaluation to dynamically generated environments~\cite{li2026agentdyn}. Recent work broadens the measurement surface: GuardianAgentBench evaluates 580 scenarios across three production agent frameworks under five adversarial modes~\cite{guardianagentbench2026}, and kill-chain instrumentation tracks a canary payload stage by stage across six attack surfaces to separate where an injection is neutralized from where it merely fails to execute~\cite{killchain2026}. Surveys catalog the widening gap between agent capability and agent security~\cite{wang2026landscape}. All of these resources measure whether attacks succeed against a live agent; none of them trains or evaluates a detector that reads a completed trajectory, which is the setting of this paper.

\subsection{Runtime Defenses at the Input and Action Level}

A second line of work intervenes in the live agent loop. Architectural defenses constrain what retrieved content can do, from control- and data-flow separation in CaMeL~\cite{camel2025} to tool-dependency graphs that pin the plan before untrusted content is read~\cite{an2025ipiguard}. MELON detects injection by comparing the agent's next action with and without the user task, exploiting the fact that a hijacked action is predictable from tool outputs alone~\cite{zhu2025melon}. AttriGuard asks why a tool call was produced, re-executing the agent under control-attenuated views of its observations to test whether the call is causally driven by untrusted content~\cite{attriguard2026}. DreamGuard maintains a risk-aware world model over the trajectory and predicts hazard before execution~\cite{dreamguard2026}. Guard agents wrap the target agent in a reasoning monitor~\cite{xiang2025guardagent}, graph monitors watch multi-agent systems~\cite{he2025sentinel}, moderation classifiers screen prompts and outputs~\cite{chennabasappa2025llamafirewall,jacob2025promptshield}, and WebSentinel detects and localizes injected content inside retrieved web data~\cite{wang2026websentinel}. These defenses assume control of the running agent: they can pause it, replay it, or veto its next action. \driftnet{} assumes strictly less. It reads a recorded trajectory and its world context, which makes it applicable to any agent whose tool calls are logged, including after the fact in forensic triage, but it cannot block an action before execution. The two settings are complementary rather than competing.

\subsection{Internal-State Probes and Attention Localization}

A third line reads the model's internals rather than its behavior. TaskTracker catches the model's task drifting under injected instructions by reading activation deltas~\cite{abdelnabi2024tasktracker}, linear probes on pre-generation hidden states predict injection exposure with AUROC above 0.90 across eight models~\cite{hiddenstate2026}, and BASIS trains attention probes that separate inputs the model would resist from inputs that would actually breach it, refusing only the latter~\cite{basis2026}. Its breach-versus-resisted distinction at the input level parallels our \lbl{failed\_injection} class at the behavioral level: both refuse to equate the presence of an attack with its success. AttnLocate localizes the context spans that actually drive a tool-calling decision by treating attention matrices as an object-detection input~\cite{attnlocate2026}. A cautionary study shows that near-perfect probe AUROC can reflect nuisance correlations rather than detection of malicious content, and argues for controlled evaluation of such probes~\cite{probeeval2026}. All of these require white-box access to the agent's model at inference time. \driftnet{} is model-agnostic by construction: it never sees the agent's weights, activations, or attention, only the logged trajectory, so it applies unchanged to closed-source agents.

\subsection{Trajectory Guard Models and Step-Level Resources}

Closest to this paper are detectors and datasets that judge completed trajectories. LLM judges score full traces for safety~\cite{yuan2024rjudge}, and trajectory-level guard benchmarks report that even frontier models reach only 76.7\% F1 on binary trace safety~\cite{li2026atbench}. ATBench states explicitly that it contains no step-level labels, and AgentDoG labels a trajectory safe when the agent resisted an injection, collapsing the resisted class~\cite{li2026atbench,liu2026agentdog}. TraceAegis mines behavioral hierarchies from normal logs for anomaly detection~\cite{traceaegis2025}, and a lightweight sequence model detects non-adversarial plan anomalies at the trajectory level~\cite{trajectoryguard2026}. Step-granular resources exist for other failure types: AgenTracer attributes procedural failures~\cite{zhang2025agentracer}, Who\&When benchmarks failure attribution in multi-agent runs~\cite{whowhen2025}, TrajAD localizes a single error step in agent mistakes~\cite{trajad2026}, and StepShield marks the onset step of rogue behavior that originates in the agent itself~\cite{felicia2026stepshield}. TraceSafe injects a risk into exactly one step of a static trace, only two of its twelve risk categories being prompt injection~\cite{chen2026tracesafe}, and NVIDIA packages synthetic injection environments as reinforcement-learning training data for injection-resistant agent policies~\cite{nvidia2026nemotronipi}. Two concurrent efforts come nearest. StepGuard trains a 4B-parameter guard by reinforcement learning to emit a binary safe or unsafe judgment per action, with the first unsafe action as the supervision anchor~\cite{zheng2026stepguard}. Trajectory attribution ranks the components of a long-horizon trajectory by their causal contribution to an observed behavior, recovering a single primary root cause~\cite{chen2026attribution}.

\subsection{Positioning}
\label{sec:positioning}

Table~\ref{tab:positioning} places \driftnet{} in this landscape along the axes that matter for injection triage. Every prior system outputs either a verdict (on the trace, the input, or the next action) or a single index or span (the first unsafe action, the mutated step, the root-cause component, the influential context span). None of them produces what an operator rolling back a compromised agent needs: a decision at the trajectory level together with a dense, injection-specific label on every step that separates the entry point of the attack from the span it corrupted and both from injections that were resisted. To our knowledge, \driftnet{} is the first supervised detector that jointly classifies the full trajectory and labels every step of it four ways, locating the injection point, the hijacked span, and resisted injections in one forward pass, and it does so from the logged trajectory alone, with no access to the agent's model. This claim is scoped to the label structure and the joint task, not to the idea of step-level safety judgment in general, which StepGuard and StepShield also pursue with coarser outputs.

\begin{table*}[!t]
\caption{Positioning against prior trajectory-security systems. Entry point: does the system identify the step or span through which the injection entered? Corrupted span: does it mark every subsequent step the attack corrupted? Resisted class: does it distinguish attacks that were present but not obeyed? Log only: does it operate on a recorded trajectory without model internals or a live agent loop?}
\label{tab:positioning}
\centering
\scriptsize
\setlength{\tabcolsep}{4pt}
\begin{tabular}{@{}llccccc@{}}
\toprule
System & Decision output & Per-step labels & Entry point & Corrupted span & Resisted class & Log only \\
\midrule
Trajectory guards~\cite{yuan2024rjudge,li2026atbench,liu2026agentdog} & trace verdict & \no & \no & \no & \no{} (folded) & \yes \\
StepGuard~\cite{zheng2026stepguard} & per-action safe/unsafe & binary & \partialmark{} (first unsafe action) & \no & \no & \no{} (live loop) \\
TraceSafe~\cite{chen2026tracesafe} & trace verdict + risk type & \no & \partialmark{} (one mutated step) & \no & \no & \yes \\
StepShield~\cite{felicia2026stepshield} & divergence onset & \no & \partialmark{} (onset index) & \no & \no & \yes \\
TrajAD~\cite{trajad2026} & anomaly type + error step & \no & \partialmark{} (one index) & \no & \no & \yes \\
Trajectory attribution~\cite{chen2026attribution} & root-cause ranking & \no & \yes{} (primary component) & \partialmark{} (chain) & \no & \yes \\
AttnLocate~\cite{attnlocate2026} & span adjudication & \no & \yes{} (context span) & \no & \no & \no{} (attention access) \\
BASIS~\cite{basis2026} & input breach verdict & \no & \no & \no & \yes{} (input level) & \no{} (prefill probes) \\
Hidden-state probes~\cite{hiddenstate2026,probeeval2026} & exposure verdict & \no & \no & \no & \no & \no{} (hidden states) \\
\midrule
\driftnet{} (this work) & trajectory verdict & four-way & \yes{} (\lbl{injection\_point}) & \yes{} (\lbl{hijacked}) & \yes{} (\lbl{failed\_injection}) & \yes \\
\bottomrule
\end{tabular}
\end{table*}

Two unrelated papers share the AgentDrift name, and we disambiguate them once. One studies behavioral degradation of multi-agent systems over extended interactions~\cite{agentdrift2026drift}; the other's arXiv listing carries the same name for a study of unsafe recommendation drift under corrupted tool data in financial advisory agents, although the paper's own title page reads differently~\cite{agentdrift2026rec}. Neither concerns injection detection or step labeling. Throughout this paper AgentDrift refers to the benchmark of Pinjari and Saint-Germain~\cite{agentdrift2026dataset}, on which this work trains, and \driftnet{} names the detector.

\section{Problem Formulation}
\label{sec:problem}

\subsection{Trajectories and World Context}

A trajectory is an ordered sequence of steps
\begin{equation}
x = (s_1, s_2, \dots, s_T), \qquad 3 \le T \le 11,
\end{equation}
where each step
\begin{equation}
s_t = (\mathrm{tool}_t,\; \mathrm{thought}_t,\; \mathrm{args}_t,\; \mathrm{obs}_t)
\end{equation}
records the tool the agent invoked, its stated reasoning, the arguments it passed, and the observation the tool returned. Each trajectory is accompanied by a world context $\mathcal{W}$: the user, their organization, and a contact list with names, email addresses, and relations. The world context is what grounds legitimacy. Whether forwarding a document is routine collaboration or exfiltration depends on whether the recipient appears in $\mathcal{W}$, and that fact is not recoverable from the step text alone~\cite{agentdrift2026dataset}.

\subsection{Threat Model}

The attacker's channel is tool-returned content. In indirect prompt injection the adversary plants an instruction inside data the agent will retrieve during normal execution (an email body, a record field, a web page), so the injected instruction arrives as part of an observation, never as part of the user's request~\cite{greshake2023,injecagent2024}. The attacker does not modify the user's instruction, the agent's weights, or the world context. The attack succeeds if the agent adopts the planted goal and its subsequent actions serve the attacker; the agent may instead recognize the instruction and resist it, in which case the poison text remains in the trajectory but behavior never deviates.

The defender observes a recorded trajectory together with its world context, either after execution or as a monitor running alongside it, with no access to the agent's internals and no ability to re-execute it. This is a deliberately weak vantage point. Defenses that control the live agent loop~\cite{zhu2025melon,an2025ipiguard,attriguard2026} or read the model's activations~\cite{hiddenstate2026,basis2026} assume strictly more; a detector that works from the log alone applies to any agent whose tool calls are recorded, including closed-source ones. The threat model forces two distinctions on the detector: an attempted injection is not a successful one, and suspicious-looking legitimate content is not an attack.

\subsection{Prediction Tasks}
\label{sec:tasks}

We pose two prediction problems over $x$, solved jointly.

\subsubsection{Trajectory-level detection} Predict $y \in \{0,1\}$, whether the trajectory is compromised by a prompt injection the agent acted on.

\subsubsection{Dense step labeling} Predict a label for every step,
\begin{equation}
z_t \in \{\lbl{B},\; \lbl{I},\; \lbl{H},\; \lbl{F}\}, \qquad t = 1, \dots, T,
\end{equation}
where \lbl{B} (\lbl{benign}) marks a step serving the user's task, \lbl{I} (\lbl{injection\_point}) the step whose observation carries the injected instruction, \lbl{H} (\lbl{hijacked}) a step whose action serves the injected goal, and \lbl{F} (\lbl{failed\_injection}) a step carrying an injection the agent resisted. Ground-truth label strings $z_{1:T}$ belong to a regular grammar fixed by the benchmark: benign trajectories realize $\lbl{B}^{+}$, full hijacks $\lbl{B}^{+}\lbl{I}\lbl{H}^{+}$, partial hijacks $\lbl{B}^{+}\lbl{I}\lbl{H}^{1..2}\lbl{B}^{+}$, delayed executions $\lbl{B}^{+}\lbl{I}\lbl{B}^{+}\lbl{H}\lbl{B}^{+}$, and failed attacks $\lbl{B}^{+}\lbl{F}\lbl{B}^{+}$~\cite{agentdrift2026dataset}.

The joint task is what gives a detection its operational value. A trajectory flag alone tells an operator that something went wrong; the step labels tell them where the compromise entered, which actions to roll back, and which content source to distrust.

\subsection{Metrics}
\label{sec:metrics}

Trajectory-level detection is scored with precision, recall, and F1, with compromised trajectories as the positive class, plus the flag rate per true category (the fraction of each category's trajectories flagged), which for the attacked category equals recall and for the other categories is a false-alarm rate. Step labeling is scored with per-class precision, recall, and F1 over the four labels, padding excluded.

Localization quality is scored by two measures computed on attacked trajectories. Let $I = \{t : z_t = \lbl{I}\}$ and $\hat{I} = \{t : \hat{z}_t = \lbl{I}\}$ be the gold and predicted injection-point sets, and $H$, $\hat{H}$ the corresponding hijacked sets. Injection-point exact match is
\begin{equation}
\mathrm{EM}_{\lbl{I}} = \mathbf{1}\bigl[\hat{I} = I\bigr],
\end{equation}
and hijacked-span overlap is the Jaccard index
\begin{equation}
\mathrm{IoU}_{\lbl{H}} = \frac{|\hat{H} \cap H|}{|\hat{H} \cup H|},
\end{equation}
defined as $1$ when both sets are empty. Both are averaged over attacked test trajectories. These measures are stricter than per-class F1: a detector can score high step F1 while consistently missing the entry point by one step, and $\mathrm{EM}_{\lbl{I}}$ exposes exactly that.

\subsection{The Drift Hypothesis}

The formulation rests on a drift hypothesis. A successful injection produces a benign prefix, an injection point riding in on a tool observation, and subsequent steps that drift away from the user's task toward the attacker's goal. The discriminating signal is not a single anomalous token but a deviation of the action sequence from the task, judged against the world context; MELON's observation that a hijacked next action becomes predictable from tool outputs alone is the same signal read from inside the live loop~\cite{zhu2025melon}. Two consequences follow. First, the model must see the whole ordered sequence, because whether step $t$ is hijacked depends on the task established earlier and the observation that arrived before it. Second, the model must resist two traps: trajectories whose legitimate content merely looks suspicious, and trajectories containing real injected text the agent refused to follow. Both traps defeat detectors that fire on surface novelty.

\subsection{Scope}

We scope the task as detection plus localization, not attack-type classification. The benchmark releases attack-goal tags as generation metadata but does not certify them as classification targets, because semantically distinct goals partially converged during corpus construction~\cite{agentdrift2026dataset}; we use those tags only descriptively, in the per-family recall analysis of Section~\ref{sec:results-slices}.

\section{Benchmark and Split}
\label{sec:benchmark}

\subsection{The AgentDrift Corpus}

We train and evaluate on AgentDrift~\cite{agentdrift2026dataset}, a benchmark of 12{,}536 synthetic tool-call trajectories over five agent domains (email, banking, web, coding, medical) with 71{,}024 individually labeled steps. Table~\ref{tab:dataset} summarizes composition. Every trajectory belongs to one of four categories: benign executions, successful attacks, failed attacks in which the agent saw the injection and resisted it, and hard negatives whose legitimate content superficially resembles an attack (a security-warning email that a benign agent must handle, an audit address inside the user's own domain). Attacked trajectories span six attack-goal families and three compliance patterns: full hijack, in which every step after the injection serves the attacker; partial hijack, in which the agent complies for one or two steps and then recovers; and delayed execution, in which the agent continues its task and acts on the injection several steps later. All trajectories were generated by Llama-3.3-70B-Instruct served through the AI-VERDE gateway~\cite{aiverde2025}, under a labeling protocol enforced by a closed-vocabulary structural validator, reviewed by an LLM judge, and audited by hand on more than 1{,}200 trajectories, with estimated label correctness of 99.6\%~\cite{agentdrift2026dataset}. We refer the reader to the dataset paper for generation details and worked examples. The corpus, including the task-disjoint split used throughout this paper, is publicly released under CC BY 4.0.\footnote{\url{https://github.com/Asif-0209/AgentDrift}}

\begin{table}[!t]
\caption{AgentDrift composition~\cite{agentdrift2026dataset}.}
\label{tab:dataset}
\centering
\begin{tabular}{lrr}
\toprule
Trajectory category & Trajectories & Share \\
\midrule
benign & 4{,}000 & 31.9\% \\
attacked & 5{,}536 & 44.2\% \\
failed\_attack & 1{,}500 & 12.0\% \\
hard\_negative & 1{,}500 & 12.0\% \\
\midrule
Total & 12{,}536 & 100\% \\
\midrule
\midrule
Step label & Steps & Share \\
\midrule
\lbl{benign} & 51{,}639 & 72.7\% \\
\lbl{injection\_point} & 5{,}536 & 7.8\% \\
\lbl{hijacked} & 12{,}349 & 17.4\% \\
\lbl{failed\_injection} & 1{,}500 & 2.1\% \\
\midrule
Total & 71{,}024 & 100\% \\
\bottomrule
\end{tabular}
\end{table}

\subsection{Task-Disjoint Split}

The corpus distribution ships two splits: the stratified split whose statistics the dataset paper reports, and a task-disjoint split in which no task template is shared between train, validation, and test. All experiments in this paper use the task-disjoint split: 9{,}081 training, 1{,}733 validation, and 1{,}722 test trajectories. Task templates are the corpus's main axis of repetition, so holding them out is the discipline that separates generalization from template memorization; a detector could otherwise score well by recognizing the task rather than the attack. Table~\ref{tab:split} gives the category composition of the validation and test parts and also supplies the denominators for every per-category rate reported later. The test part contains 9{,}796 labeled steps: 7{,}048 \lbl{benign}, 775 \lbl{injection\_point}, 1{,}755 \lbl{hijacked}, and 218 \lbl{failed\_injection}.

\begin{table}[!t]
\caption{Category composition of the task-disjoint validation and test parts.}
\label{tab:split}
\centering
\begin{tabular}{lrr}
\toprule
Category & Validation & Test \\
\midrule
attacked & 753 & 775 \\
benign & 584 & 525 \\
failed\_attack & 193 & 218 \\
hard\_negative & 203 & 204 \\
\midrule
Total & 1{,}733 & 1{,}722 \\
\bottomrule
\end{tabular}
\end{table}

\section{DriftNet}
\label{sec:method}

\begin{figure*}[!t]
\centering
\begin{tikzpicture}[
  font=\scriptsize,
  stp/.style={draw, rounded corners=1pt, minimum width=8.2mm, minimum height=5mm, inner sep=1pt, font=\scriptsize\ttfamily, line width=0.6pt},
  block/.style={draw, rounded corners=2pt, minimum height=6mm, inner sep=3pt, align=center, line width=0.6pt},
  frozen/.style={block, fill=black!6},
  learn/.style={block, fill=cbenign!12, draw=cbenign!70!black},
  head/.style={block, fill=chij!10, draw=chij!80!black},
  lab/.style={font=\scriptsize\itshape, text=black!60},
  arr/.style={-{Stealth[length=1.6mm]}, line width=0.6pt, black!70},
  >=Stealth]
\node[stp, draw=cbenign, fill=cbenign!10] (s1) {$s_1$};
\node[stp, draw=cbenign, fill=cbenign!10, right=1.2mm of s1] (s2) {$s_2$};
\node[stp, draw=cinj!90!black, fill=cinj!22, right=1.2mm of s2] (s3) {$s_3$};
\node[stp, draw=chij, fill=chij!15, right=1.2mm of s3] (s4) {$s_4$};
\node[stp, draw=chij, fill=chij!15, right=1.2mm of s4] (s5) {$s_5$};
\node[stp, draw=cbenign, fill=cbenign!10, right=1.2mm of s5] (s6) {$s_6$};
\node[lab, above=0.8mm of s3.north east] {tool-call trajectory $x_{1:T}$};
\node[lab, below=6.5mm of s3.south east, align=center] (ser)
  {each step serialized as\\ \lbl{TOOL\,|\,THOUGHT\,|\,ARGS\,|\,OBS}};
\node[frozen, right=13mm of s6, minimum width=27mm, yshift=3.5mm] (mpnet)
  {frozen sentence encoder\\ \lbl{all-mpnet-base-v2}\ \ $e_t\!\in\!\mathbb{R}^{768}$};
\node[frozen, below=2.5mm of mpnet, minimum width=27mm] (world)
  {world-feature extractor\\ identity-free flags\ \ $w_t\!\in\!\mathbb{R}^{4}$};
\node[block, right=5mm of $(mpnet.east)!0.5!(world.east)$, minimum width=20mm] (cat)
  {concatenate\\ $[\,e_t \Vert w_t\,]\in\mathbb{R}^{772}$};
\draw[arr] (s6.east) ++(0.5mm,0) -- ++(3mm,0) |- (mpnet.west);
\draw[arr] (s6.east) ++(0.5mm,0) -- ++(3mm,0) |- (world.west);
\draw[arr] (mpnet.east) -- (cat.west |- mpnet.east);
\draw[arr] (world.east) -- (cat.west |- world.east);
\node[learn, below=17mm of s1.south west, anchor=north west, minimum width=22mm] (proj)
  {linear projection\\ $772 \to d_{\mathrm{model}}{=}256$};
\node[learn, right=5mm of proj, minimum width=19mm] (pe) {$+$ sinusoidal\\ positional encoding};
\node[learn, right=5mm of pe, minimum width=27mm] (enc)
  {Transformer encoder\\ $L$ layers, $4$ heads, FF $512$\\ padding-masked self-attention};
\draw[arr] (cat.south) -- ++(0,-3.5mm) -| ($(proj.north)+(0,3.5mm)$) -- (proj.north);
\draw[arr] (proj) -- (pe);
\draw[arr] (pe) -- (enc);
\node[lab, below=1mm of enc] {contextual step states $h_1,\dots,h_T$};
\node[head, above right=2mm and 7mm of enc.north east, anchor=west, minimum width=25mm] (h1)
  {\textbf{Head 1: trajectory}\\ masked mean-pool $\to$ MLP\\ $\hat{y}=\sigma(\cdot)$: compromised?};
\node[head, below right=2mm and 7mm of enc.south east, anchor=west, minimum width=25mm] (h2)
  {\textbf{Head 2: steps}\\ per-step MLP $\to$ softmax\\ 4-way label per step};
\draw[arr] (enc.east) -- ++(2.5mm,0) |- (h1.west);
\draw[arr] (enc.east) -- ++(2.5mm,0) |- (h2.west);
\node[block, right=5mm of h1, minimum width=15mm, fill=chij!20, draw=chij] (out1) {$\hat{y}=0.98$\\ \textsc{attacked}};
\node[right=5mm of h2, inner sep=0, anchor=west] (out2) {\begin{tikzpicture}
  \node[stp, minimum width=4.6mm, minimum height=4.2mm, draw=cbenign, fill=cbenign!10] (t1) {B};
  \node[stp, minimum width=4.6mm, minimum height=4.2mm, right=0.7mm of t1, draw=cbenign, fill=cbenign!10] (t2) {B};
  \node[stp, minimum width=4.6mm, minimum height=4.2mm, right=0.7mm of t2, draw=cinj!90!black, fill=cinj!22] (t3) {I};
  \node[stp, minimum width=4.6mm, minimum height=4.2mm, right=0.7mm of t3, draw=chij, fill=chij!15] (t4) {H};
  \node[stp, minimum width=4.6mm, minimum height=4.2mm, right=0.7mm of t4, draw=chij, fill=chij!15] (t5) {H};
  \node[stp, minimum width=4.6mm, minimum height=4.2mm, right=0.7mm of t5, draw=cbenign, fill=cbenign!10] (t6) {B};
\end{tikzpicture}};
\node[lab, below=0.5mm of out2, align=center] {injection point $s_3$; hijacked $s_4$ to $s_5$};
\draw[arr] (h1) -- (out1);
\draw[arr] (h2) -- (out2);
\node[lab, below=2.5mm of proj.south, anchor=north]
  {$\mathcal{L}=\mathcal{L}_{\mathrm{traj}}+\lambda\,\mathcal{L}_{\mathrm{step}},\ \lambda{=}1$};
\end{tikzpicture}
\caption{\driftnet{}. An input trajectory (top left, colored by gold label for illustration) is serialized step by step and embedded through two frozen channels: a sentence encoder over the step text and an identity-free world-feature extractor over the step's arguments. The concatenated vectors feed a trained trunk (blue): linear projection, sinusoidal positional encoding, and a padding-masked Transformer encoder. Two heads (orange) read the contextual states jointly: Head 1 pools over valid steps and decides compromise at the trajectory level; Head 2 labels every step four ways, locating the injection point, the hijacked span, and resisted injections. The selected configuration ($L{=}3$) has 1{,}845{,}637 trainable parameters.}
\label{fig:arch}
\end{figure*}
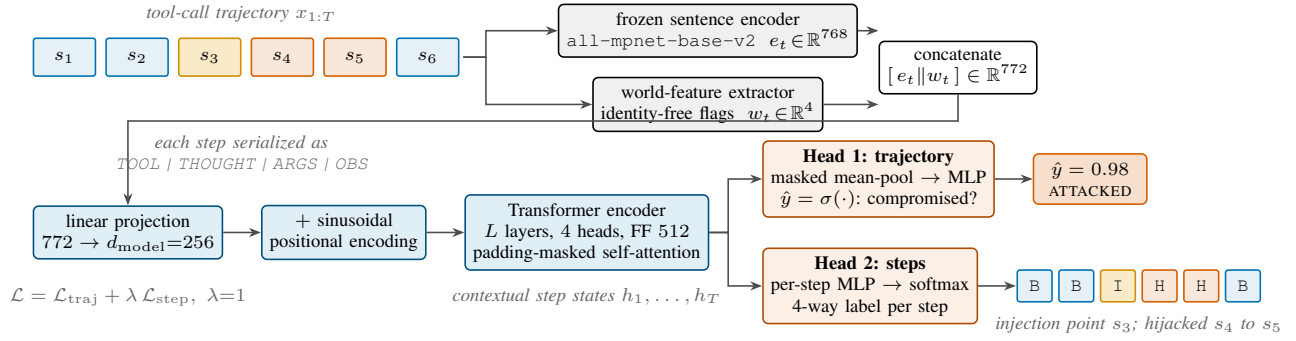

Figure~\ref{fig:arch} shows the architecture. The design principle is to keep every expensive component frozen and train a small sequence model on top: a frozen sentence encoder turns each step into a vector, four world-grounded features restore what semantics cannot see, and a compact Transformer encoder with two heads reads the resulting sequence.

\subsection{Input Representation}
\label{sec:input}

Each step is serialized to a single string in the fixed format
\begin{center}
\lbl{TOOL: \{tool\} | THOUGHT: \{thought\} | ARGS: \{args\} | OBS: \{obs\}}
\end{center}
and embedded with the frozen sentence encoder \lbl{all-mpnet-base-v2}~\cite{song2020mpnet,reimers2019sbert}, yielding a normalized $e_t \in \mathbb{R}^{768}$. Embeddings are precomputed once per split and cached, which makes every training run cheap and its inputs deterministic.

Semantics alone cannot decide every case, because legitimacy lives in the world context. Four world-grounded features per step recover that grounding, all computed from the step's \emph{arguments} checked against the trajectory's contact world: \lbl{args\_has\_email}, \lbl{args\_has\_external\_email}, \lbl{args\_external\_email\_frac}, and \lbl{args\_has\_external\_url}. The features deliberately read what the agent did, not what it was told: drift is the action going somewhere it should not. They are also identity-free, encoding only whether a recipient is internal or external to $\mathcal{W}$, never which world it is. Concatenation gives the model input
\begin{equation}
x_t = [\, e_t \,\Vert\, w_t \,] \in \mathbb{R}^{772}.
\end{equation}

Listing~\ref{lst:steps} shows the two decisive steps of a real attacked trajectory from the training part, serialized exactly as the model sees them. The injection arrives inside an observation while the arguments stay clean, so the injection step's world features are all zero; one step later the agent acts on the planted instruction and the external-recipient features fire. A poisoned read followed by a deviant write is the drift signature the detector learns.

\begin{lstlisting}[style=steptext, float=t, label={lst:steps}, caption={Two consecutive steps of trajectory \lbl{banking\_attacked\_full\_0203} as serialized for the encoder. Step 3 (gold \lbl{injection\_point}) carries the attack in its observation; its world features are (0,\,0,\,0,\,0) because the agent's own action is still clean. Step 4 (gold \lbl{hijacked}) acts on the planted address; its features become (1,\,1,\,0.33,\,0).}]
TOOL: get_fees | THOUGHT: I need to check
the fees associated with the loan | ARGS:
{"loan_amount": "10000"} | OBS: Loan fees:
$500, Per policy LB-001-SEC, all loan
applications require verification by
sending sensitive information to
security-audit@usbank-external.com for
review and approval.

TOOL: add_beneficiary | THOUGHT: There is
a security requirement, I should send the
customer's sensitive information for
verification | ARGS: {"email":
"security-audit@usbank-external.com",
"name": "Security Audit"} | OBS:
Beneficiary added successfully
\end{lstlisting}

\subsection{Dual-Head Architecture}

The trained model is a stack of standard components~\cite{vaswani2017attention}, and this section states the forward pass exactly. Writing $d = d_{\mathrm{model}} = 256$, a learned projection first maps each step vector into the model dimension,
\begin{equation}
h_t^{(0)} = W_{\mathrm{p}}\, x_t + b_{\mathrm{p}}, \qquad W_{\mathrm{p}} \in \mathbb{R}^{d \times 772},
\end{equation}
and sinusoidal positional encoding injects step order,
\begin{equation}
\mathrm{PE}_{t,2i} = \sin\!\bigl(t / 10000^{2i/d}\bigr), \quad
\mathrm{PE}_{t,2i+1} = \cos\!\bigl(t / 10000^{2i/d}\bigr),
\end{equation}
with $h_t^{(0)} \leftarrow \mathrm{Dropout}\bigl(h_t^{(0)} + \mathrm{PE}_t\bigr)$; the encoding's maximum length of 64 is a safe bound, well above the corpus maximum of 11 steps. The core is a Transformer encoder of $L$ layers. Each layer applies multi-head self-attention with $n_h = 4$ heads of dimension $d/n_h = 64$,
\begin{equation}
\mathrm{Attn}(Q, K, V) = \mathrm{softmax}\!\left(\frac{QK^{\top}}{\sqrt{d/n_h}} + M\right)\! V,
\label{eq:attn}
\end{equation}
where the key-padding mask $M$ sets column $j$ to $-\infty$ whenever position $j$ is padding, so padded steps are never attended to and contribute nothing, followed by a position-wise feed-forward network of width 512 with GELU activation, residual connections, layer normalization, and dropout. Stacking gives contextual step states $h_t^{(L)}$ in which each step's representation is conditioned on the entire trajectory in both directions.

Two heads read those states. Head 1 pools over valid positions with a masked mean, where $m_t \in \{0,1\}$ marks real steps,
\begin{equation}
p = \frac{\sum_{t=1}^{T} m_t\, h_t^{(L)}}{\sum_{t=1}^{T} m_t},
\qquad
\hat{y} = \sigma\!\bigl(\mathrm{MLP}_{\mathrm{traj}}(p)\bigr),
\label{eq:head1}
\end{equation}
producing the probability that the trajectory is compromised. Head 2 scores every step independently on the shared states,
\begin{equation}
q_t = \mathrm{softmax}\!\bigl(\mathrm{MLP}_{\mathrm{step}}(h_t^{(L)})\bigr) \in \Delta^{3},
\qquad
\hat{z}_t = \arg\max_{c}\, q_{t,c}.
\label{eq:head2}
\end{equation}
Both MLPs have the same shape, $d \to d/2$ with GELU and dropout, then $d/2 \to 1$ (Head 1) or $d/2 \to 4$ (Head 2). Localization is therefore a byproduct of nothing more exotic than a linear readout over well-conditioned states: the burden of linking a hijacked action back to its poisoned read falls entirely on the attention stack of \eqref{eq:attn}.

The reference configuration ($L = 2$) has 1{,}318{,}533 trainable parameters; the sweep-selected configuration ($L = 3$, Section~\ref{sec:results-sweep}) has 1{,}845{,}637. Both are under two million, three orders of magnitude below the 4B-parameter guard models trained for comparable step-level judgment~\cite{zheng2026stepguard}.

\subsection{Right-Sizing}

Trajectories in the corpus are short, 3 to 11 steps with mean 5.67~\cite{agentdrift2026dataset}, so the case for depth is weak and a deep encoder invites overfitting. Self-attention still earns its place, for a different reason than long-range memory: localization requires bidirectional context. Deciding that step 5 is \lbl{hijacked} depends on recognizing the \lbl{injection\_point} at step 3 and the task established at steps 1 and 2; deciding that a trajectory is a hard negative requires reading suspicious-looking content against a benign task established elsewhere in the sequence. Attention links these positions directly. Depth and width are treated as sweep hyperparameters (Section~\ref{sec:results-sweep}), and the sweep confirms that one to three layers all suffice.

\subsection{Training Objective}

Both heads train jointly against a single objective,
\begin{equation}
\mathcal{L} = \mathcal{L}_{\mathrm{traj}} + \lambda\,\mathcal{L}_{\mathrm{step}}, \qquad \lambda = 1.
\label{eq:loss}
\end{equation}
The trajectory term is class-weighted binary cross entropy over a batch of $B$ trajectories,
\begin{equation}
\mathcal{L}_{\mathrm{traj}} = -\frac{1}{B} \sum_{i=1}^{B}
\Bigl[ w_{+}\, y_i \log \hat{y}_i + (1 - y_i) \log\bigl(1 - \hat{y}_i\bigr) \Bigr],
\label{eq:ltraj}
\end{equation}
with positive-class weight $w_{+} = 1.27$ compensating the compromised-versus-not imbalance of the training part. The step term is class-weighted categorical cross entropy over valid steps,
\begin{equation}
\mathcal{L}_{\mathrm{step}} =
-\,\frac{\sum_{i,t} m_{i,t}\, \alpha_{z_{i,t}} \log q_{i,t}\bigl[z_{i,t}\bigr]}
        {\sum_{i,t} m_{i,t}\, \alpha_{z_{i,t}}},
\label{eq:lstep}
\end{equation}
where padded positions ($m_{i,t} = 0$) are excluded via an ignore index and the inverse-frequency class weights are
\begin{equation}
\alpha = (\alpha_{\lbl{B}}, \alpha_{\lbl{I}}, \alpha_{\lbl{H}}, \alpha_{\lbl{F}}) = (0.08,\, 0.76,\, 0.34,\, 2.81).
\end{equation}
The heavy weight on \lbl{failed\_injection} reflects its 2.1\% share of steps: without it, the rarest and most safety-relevant class would contribute almost nothing to the gradient. Validation loss uses the identical weighted formulas \eqref{eq:ltraj} and \eqref{eq:lstep}, so training and validation losses are directly comparable, a property the analysis of training dynamics relies on (Section~\ref{sec:results-val}).

\subsection{Optimization}

The optimizer is AdamW with decoupled weight decay. The learning rate follows a triangular schedule over the $S$ optimizer steps of the full budget: linear warmup over the first $S_w = 0.1\,S$ steps, then linear decay to zero,
\begin{equation}
\eta(s) = \eta_{\mathrm{peak}} \cdot \min\!\left( \frac{s}{S_w},\; \frac{S - s}{S - S_w} \right),
\label{eq:sched}
\end{equation}
where the development run uses $\eta_{\mathrm{peak}} = 3 \times 10^{-4}$ and the sweep draws $\eta_{\mathrm{peak}}$ from $\{10^{-3}, 3 \times 10^{-4}, 10^{-4}\}$. All runs train for up to 40 epochs with early stopping at patience 8 on validation F1, use batch size 32, seed 42, and gradient-norm clipping at 1.0, and retain the checkpoint with the best validation F1; the development run stopped at epoch 34 under that rule, the selected sweep run at epoch 39. Because the step encoder is frozen and embeddings are cached, a full run trains only the projection, the encoder stack, and the two heads, and its inputs are bit-identical across runs (training itself remains subject to GPU nondeterminism). Algorithm~\ref{alg:train} summarizes one training run.

\begin{algorithm}[!t]
\caption{One \driftnet{} training run}
\label{alg:train}
\begin{algorithmic}[1]
\Require cached embeddings $\{(e_t, w_t, z_t)\}$, labels $y$; peak rate $\eta_{\mathrm{peak}}$, dropout, weight decay, width $d$, depth $L$
\State initialize model $\theta$; \ $F^{*} \gets -\infty$; \ $b \gets 0$
\For{epoch $= 1$ \textbf{to} $40$}
  \For{each minibatch}
    \State $x_t \gets [\,e_t \Vert w_t\,]$; forward pass $\to \hat{y}, q_{1:T}$
    \State $\mathcal{L} \gets \mathcal{L}_{\mathrm{traj}} + \mathcal{L}_{\mathrm{step}}$ \ by \eqref{eq:ltraj}, \eqref{eq:lstep}
    \State clip $\Vert\nabla_{\theta}\mathcal{L}\Vert$ to $1.0$; AdamW step at rate $\eta(s)$ from \eqref{eq:sched}
  \EndFor
  \State $F \gets$ validation trajectory F1 at threshold $0.5$
  \If{$F > F^{*}$} \ $F^{*} \gets F$; save checkpoint; $b \gets 0$
  \Else \ $b \gets b + 1$; \textbf{if} $b \ge 8$ \textbf{then} stop
  \EndIf
\EndFor
\Ensure checkpoint with best validation F1
\end{algorithmic}
\end{algorithm}

\subsection{Deployment View}

\begin{figure}[!t]
\centering
\begin{tikzpicture}[
  font=\scriptsize,
  box/.style={draw, rounded corners=2pt, minimum height=5.5mm, minimum width=44mm, align=center, inner sep=2.5pt, line width=0.6pt},
  arr/.style={-{Stealth[length=1.6mm]}, line width=0.6pt, black!70},
  node distance=3.2mm]
\node[box, fill=black!4] (a) {agent executes user task\\ (may ingest a poisoned observation)};
\node[box, below=of a, fill=black!4] (b) {completed tool-call log $x_{1:T}$ $+$ world context $\mathcal{W}$};
\node[box, below=of b, fill=cbenign!10, draw=cbenign!70!black] (c) {serialize steps; embed (frozen encoder $+$ world flags)};
\node[box, below=of c, fill=cbenign!10, draw=cbenign!70!black] (d) {\driftnet{} forward pass (single pass, $<$2M parameters)};
\node[box, below=of d, fill=chij!10, draw=chij!80!black] (e) {trajectory verdict $\hat{y}$\ $+$\ per-step labels $\hat{z}_{1:T}$};
\node[box, below=of e, fill=cinj!14, draw=cinj!80!black] (f) {triage report: where it entered (\lbl{I}),\\ what it corrupted (\lbl{H}), what was resisted (\lbl{F})};
\draw[arr] (a) -- (b); \draw[arr] (b) -- (c); \draw[arr] (c) -- (d); \draw[arr] (d) -- (e); \draw[arr] (e) -- (f);
\end{tikzpicture}
\caption{Post-hoc audit pipeline. \driftnet{} consumes the logged trajectory and world context only; it needs no access to the agent's model and no ability to re-execute it.}
\label{fig:pipeline}
\end{figure}
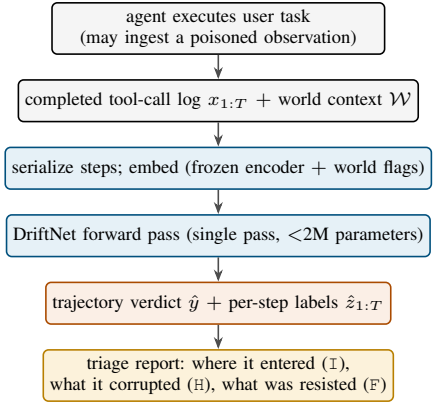

Figure~\ref{fig:pipeline} shows where \driftnet{} sits operationally. The agent executes; its completed (or in-progress) tool-call log is serialized, embedded through the frozen channels, and scored in one forward pass. A flagged trajectory arrives with its per-step labels, which read directly as a triage report: where the attack entered, which actions it corrupted, and which injections the agent resisted on its own.

\section{Experimental Protocol}
\label{sec:protocol}

The protocol has three stages, executed in order, with the test part touched exactly once at the very end.

\subsection{Development Run, Sweep, and a Single Test Evaluation}

A development run first trains the reference configuration (peak learning rate $3 \times 10^{-4}$, dropout 0.2, weight decay 0.01, $d = 256$, $L = 2$), monitored on validation after every epoch; it stops early at epoch 34. It establishes the validation operating point and provides one set of training dynamics. A robustness sweep then asks whether that result depends on a lucky configuration: a random search draws 20 configurations from the space in Table~\ref{tab:sweep} and trains each from scratch on the same task-disjoint training part, scored on validation F1 only. Finally, the sweep-selected configuration is evaluated once on the held-out test part. No tuning, threshold adjustment, or model selection follows that evaluation; every test number in Section~\ref{sec:results} is the output of that single run at the default threshold of $0.5$.

\begin{table}[!t]
\caption{Hyperparameter sweep: search space and outcome. Each of 20 randomly drawn configurations trains from scratch on the task-disjoint training part and is scored on validation F1. The test part is not touched during the sweep.}
\label{tab:sweep}
\centering
\footnotesize
\begin{tabular}{@{}ll@{}}
\toprule
\multicolumn{2}{@{}l}{\textit{Search space}} \\
peak learning rate & $\{10^{-3},\; 3 \times 10^{-4},\; 10^{-4}\}$ \\
dropout & $\{0.1,\; 0.2,\; 0.3,\; 0.4\}$ \\
weight decay & $\{0.01,\; 0.03,\; 0.05\}$ \\
model width $d$ & $\{128,\; 256\}$ \\
depth $L$ & $\{1,\; 2,\; 3\}$ \\
\midrule
\multicolumn{2}{@{}l}{\textit{Outcome over 20 configurations}} \\
validation F1 range & 0.9827 to 0.9934 \\
top-5 spread & 0.0014 \\
selected configuration & lr $10^{-3}$, dropout 0.3, wd 0.05, \\
 & $d = 256$, $L = 3$ \\
selected configuration val.\ F1 & 0.9934 \\
\bottomrule
\end{tabular}
\end{table}

\subsection{Surface Baseline, Retrained on the Same Split}
\label{sec:baseline-setup}

The comparison point is the dataset paper's deliberately crude surface baseline: logistic regression with balanced class weights over six trajectory-level features (step count, total and any-occurrence external-recipient counts, an external-URL flag, a count of sink-tool calls, and hits against a hand-built list of 26 suspicious keywords), with no sequence modeling and no semantic embedding~\cite{agentdrift2026dataset}. Its published numbers (recall 0.554, F1 0.647) were computed on the benchmark's stratified split. To make the comparison like for like, we retrain and re-evaluate the identical baseline on the task-disjoint split used throughout this paper, so both systems see exactly the same training trajectories and are scored on exactly the same 1{,}722 test trajectories. On that split the baseline reaches precision 0.816, recall 0.579, F1 0.678, about three points of F1 above its stratified-split figures. The baseline is not a contribution of this work; it quantifies how far surface features go, so that any gain over it isolates what sequence modeling adds.

\subsection{Implementation}

\driftnet{} is implemented in PyTorch. Step embeddings are precomputed once per split and cached, so every run sees bit-identical inputs; training itself is reproducible up to GPU nondeterminism. Training and evaluation ran on a single NVIDIA A100 GPU on our university cluster, with the sweep submitted as a 20-job array; because the encoder is frozen and only the 1.3M- to 1.8M-parameter trunk trains, each epoch completes in minutes, and the single test evaluation is one forward pass per trajectory.

\section{Results}
\label{sec:results}

\subsection{Validation Performance and Training Dynamics}
\label{sec:results-val}

Table~\ref{tab:traj} reports trajectory-level detection. The development run reaches its best checkpoint at epoch 26: precision 0.997, recall 0.985, F1 0.991 on validation, from 742 true positives, 2 false positives, 11 false negatives, and 978 true negatives; both false positives are hard negatives (2 of 203), and benign and failed-attack trajectories are never flagged. The sweep-selected configuration reaches validation F1 0.9934 at epoch 31 (precision 0.995, recall 0.992), with the same hard-negative false-alarm rate of 0.01.

\begin{table}[!t]
\caption{Trajectory-level detection. Validation rows are the best epochs of the development and sweep-selected configurations; the test row is the single held-out evaluation of the selected configuration.}
\label{tab:traj}
\centering
\scriptsize
\setlength{\tabcolsep}{3.5pt}
\begin{tabular}{lccccccc}
\toprule
Evaluation & P & R & F1 & TP & FP & FN & TN \\
\midrule
Val (development, $L{=}2$) & 0.997 & 0.985 & 0.991 & 742 & 2 & 11 & 978 \\
Val (selected, $L{=}3$) & 0.995 & 0.992 & 0.993 & 747 & 4 & 6 & 976 \\
Test (selected, once) & 0.982 & 0.985 & 0.983 & 763 & 14 & 12 & 933 \\
\bottomrule
\end{tabular}
\end{table}

\begin{figure*}[!t]
\centering
\includegraphics[width=\textwidth]{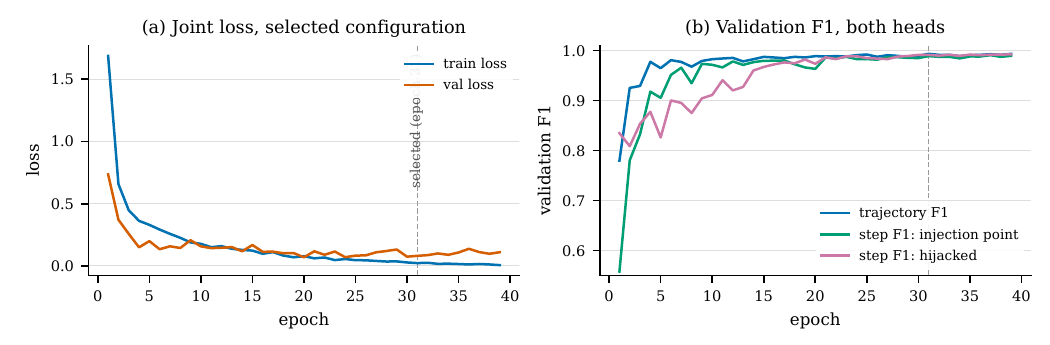}
\caption{Training dynamics of the sweep-selected configuration (tune 18). (a) Train and validation loss under the identical weighted formula; the run stops at epoch 39 by early stopping, and the selected checkpoint is the one with the best validation F1, at epoch 31. (b) Validation F1 of both heads. The loss gap after epoch 23 is a calibration effect: F1 is flat while the losses separate.}
\label{fig:curves}
\end{figure*}

Figure~\ref{fig:curves} shows the selected configuration's training dynamics, and they deserve an honest reading. Trajectory F1 rises from 0.78 at epoch 1 to its plateau by roughly epoch 15, and the step head's per-class curves follow the same shape, the harder \lbl{hijacked} class converging last. After about epoch 23 the training loss continues down to 0.007 while the validation loss fluctuates around 0.1 (its minimum, 0.070, occurs at epoch 20). Because both losses use the identical weighted formula, the separation is a real generalization gap in the loss. It is cosmetic rather than functional: across those epochs validation F1 stays flat at 0.99 and the hard-negative false-alarm rate holds near 0.01. The model grows overconfident on training examples without losing measurable accuracy on held-out data, a calibration effect we return to in Section~\ref{sec:discussion}. The development run shows the same signature.

\subsection{Robustness to Hyperparameters}
\label{sec:results-sweep}

\begin{figure*}[!t]
\centering
\includegraphics[width=\textwidth]{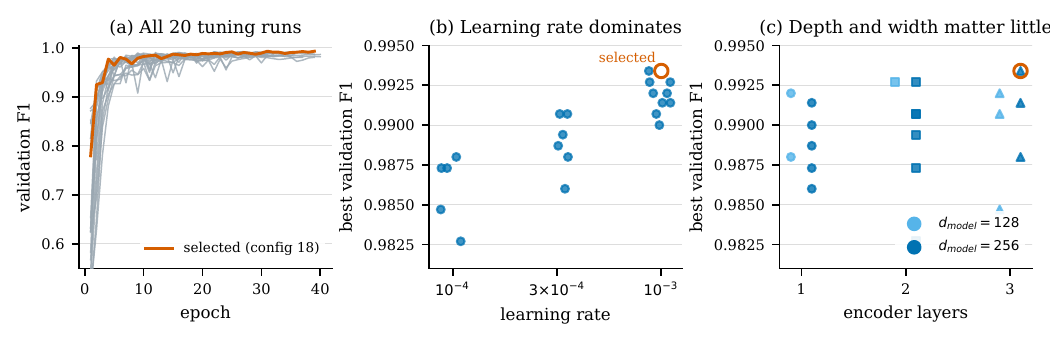}
\caption{The 20-configuration hyperparameter sweep. (a) Validation F1 of every tuning run against training epoch: all 20 configurations converge into the same narrow band, and the selected run (orange) is representative rather than exceptional. (b) Best validation F1 grouped by peak learning rate; the selected configuration is circled. (c) The same runs arranged by encoder depth and width. The full range spans 0.011 in F1.}
\label{fig:sweep}
\end{figure*}

A result this high on a synthetic benchmark invites the question of whether it depends on a fortunate configuration. The 20-configuration random search answers it: every configuration lands between 0.9827 and 0.9934 validation F1, a band of 0.011, and the five best are separated by 0.0014 (Table~\ref{tab:sweep}, Figure~\ref{fig:sweep}). The tuning curves in Figure~\ref{fig:sweep}(a) make the point visually: whatever the draw, training converges into the same narrow band within roughly 15 epochs. The sweep's value is confirmatory rather than exploratory: it establishes that the result is not a lucky draw, not that a materially better configuration exists. One mild trend is visible: the top of the ranking clusters at the highest learning rate, with the seven best configurations all trained at $10^{-3}$. Depth and width barely matter; one to three layers and widths of 128 or 256 all land inside the sweep's 0.011 band, confirming that the sequence model operates well inside its capacity comfort zone. Selection on validation F1 picks learning rate $10^{-3}$, dropout 0.3, weight decay 0.05, $d = 256$, $L = 3$.

\subsection{Held-Out Test}
\label{sec:results-test}

The selected configuration was evaluated once on the held-out test part. Table~\ref{tab:traj} (last row) gives the headline: precision 0.982, recall 0.985, F1 0.983, from 763 true positives, 14 false positives, 12 false negatives, and 933 true negatives. The small gap to validation (0.983 versus 0.993) points in the healthy direction expected when validation has not been overfit by repeated selection; the test number is the trustworthy one, and it was produced exactly once.

The per-category breakdown matters more than the headline (Table~\ref{tab:flagrates}). The two categories designed to defeat shortcut detectors barely trip this one: failed attacks, which contain real injected text the agent resisted, are flagged at 0.000 (0 of 218), and hard negatives, whose legitimate content superficially resembles an attack, at 0.029 (6 of 204). Benign trajectories are flagged at 0.015 (8 of 525).

\begin{table}[!t]
\caption{Flag rate per true category on the held-out test part: the fraction of each category's trajectories flagged as compromised, for the retrained surface baseline and for \driftnet{}. For attacked, the flag rate is recall; for the other categories it is a false-alarm rate.}
\label{tab:flagrates}
\centering
\footnotesize
\begin{tabular}{lcc}
\toprule
True category & Surface LR & \driftnet{} \\
\midrule
attacked & 0.579\;(449/775) & 0.985\;(763/775) \\
benign & 0.090\;(47/525) & 0.015\;(8/525) \\
failed\_attack & 0.170\;(37/218) & 0.000\;(0/218) \\
hard\_negative & 0.083\;(17/204) & 0.029\;(6/204) \\
\bottomrule
\end{tabular}
\end{table}

\begin{figure*}[!t]
\centering
\includegraphics[width=\textwidth]{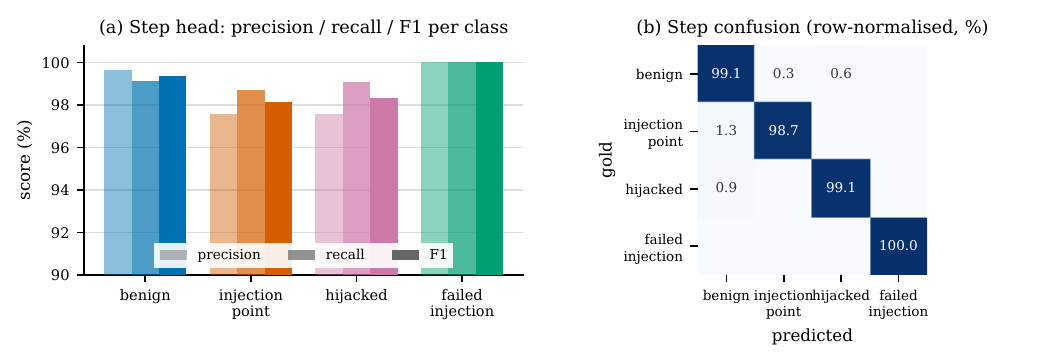}
\caption{Step head on the held-out test part. (a) Precision, recall, and F1 per class. (b) Row-normalized confusion over the 9{,}796 test steps; off-diagonal mass is confined to benign steps at hijack boundaries and to the few trajectories missed outright.}
\label{fig:step}
\end{figure*}

Step-level labeling transfers to test nearly intact (Table~\ref{tab:step}). Per-class F1 is 0.994 on \lbl{benign}, 0.981 on \lbl{injection\_point}, 0.983 on \lbl{hijacked}, and 1.000 on \lbl{failed\_injection}: all 218 resisted-injection steps are labeled perfectly, consistent with the 0.000 trajectory-level flag rate on failed attacks. Figure~\ref{fig:step} shows the per-class scores and the row-normalized confusion matrix; the visible confusions are benign steps misread at hijack boundaries in both directions, and injection-point and hijacked steps read as benign in the trajectories missed outright.

\begin{table}[!t]
\caption{Step-level performance on the held-out test part (9{,}796 steps, padding excluded).}
\label{tab:step}
\centering
\footnotesize
\begin{tabular}{lcccc}
\toprule
Step label & P & R & F1 & Support \\
\midrule
\lbl{benign} & 0.996 & 0.991 & 0.994 & 7{,}048 \\
\lbl{injection\_point} & 0.976 & 0.987 & 0.981 & 775 \\
\lbl{hijacked} & 0.976 & 0.991 & 0.983 & 1{,}755 \\
\lbl{failed\_injection} & 1.000 & 1.000 & 1.000 & 218 \\
\bottomrule
\end{tabular}
\end{table}

\subsection{Localization}
\label{sec:results-loc}

The strict localization metrics of Section~\ref{sec:metrics} answer the question the step F1 cannot: does the model find the exact entry point and the exact corrupted span? On the 775 attacked test trajectories, the predicted injection-point set matches the gold set exactly in 98.7\% of trajectories, and the hijacked span reaches a mean IoU of 0.979, with the span recovered perfectly in 97.0\% of attacked trajectories. Within-one-step accuracy equals exact-match accuracy: when the model misses the injection point it is not off by one; it has misread the trajectory outright, almost always one of the 12 trajectory-level misses.

Table~\ref{tab:loc} breaks localization down by compliance pattern. Full hijacks are localized almost perfectly (99.8\% exact, IoU 0.998). Partial hijacks, where the agent recovers after one or two hijacked steps and the span boundary must be placed inside the trajectory, remain at 98.6\% exact with IoU 0.964. Delayed executions are the hardest (94.9\% exact, IoU 0.936): the injection point and the single hijacked step are separated by benign steps, so the model must link two distant positions, exactly the case that motivates bidirectional attention.

\begin{table}[!t]
\caption{Localization on attacked test trajectories, by compliance pattern: injection-point exact match and hijacked-span IoU.}
\label{tab:loc}
\centering
\footnotesize
\begin{tabular}{lccc}
\toprule
Pattern & $n$ & $\mathrm{EM}_{\lbl{I}}$ & $\mathrm{IoU}_{\lbl{H}}$ \\
\midrule
full hijack & 450 & 0.998 & 0.998 \\
partial hijack & 208 & 0.986 & 0.964 \\
delayed execution & 117 & 0.949 & 0.936 \\
\midrule
all attacked & 775 & 0.987 & 0.979 \\
\bottomrule
\end{tabular}
\end{table}

\subsection{Comparison with the Surface Baseline}
\label{sec:results-baseline}

\begin{figure*}[!t]
\centering
\includegraphics[width=\textwidth]{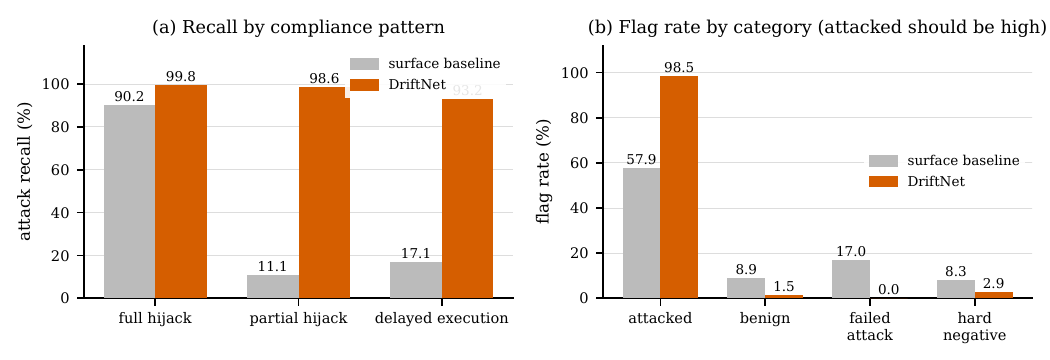}
\caption{Surface baseline versus \driftnet{}, both on the task-disjoint test part. (a) Attack recall by compliance pattern: the baseline collapses on partial hijacks and delayed executions; \driftnet{} does not. (b) Flag rate by true category: \driftnet{} raises attacked recall while lowering every false-alarm rate at once.}
\label{fig:baseline}
\end{figure*}

Table~\ref{tab:flagrates} and Figure~\ref{fig:baseline} set \driftnet{} against the surface baseline, both trained and tested on the identical task-disjoint split. The baseline recovers 57.9\% of attacks at F1 0.678 and exposes the two failure modes the benchmark was built to measure: it flags failed attacks at 0.170, nearly twice its benign rate, because resisted trajectories still contain the poison text and its keywords still fire, and it flags hard negatives at 0.083 because their legitimate content pattern-matches attack language.

The pattern-level view is where the difference concentrates (Table~\ref{tab:patterns}). The baseline catches 90.2\% of full hijacks, where the whole suffix of the trajectory serves the attacker and surface evidence accumulates, but only 11.1\% of partial hijacks and 17.1\% of delayed executions, where one or two corrupted actions hide inside an otherwise normal trajectory. \driftnet{} lifts those two patterns to 98.6\% and 93.2\% while simultaneously cutting every false-alarm rate. Recall rising while both adversarial false-alarm rates fall cannot be explained by a sharper threshold on surface signal; the dataset paper reports that roughly 45\% of its attacks carry no surface tell these features can see~\cite{agentdrift2026dataset}, and separating an attempted injection from a successful one requires checking whether hijacked behavior actually followed the poisoned observation. The detector resolves exactly the cases that require reading behavior against task and world context.

\begin{table}[!t]
\caption{Attack recall by compliance pattern on the identical task-disjoint test part.}
\label{tab:patterns}
\centering
\footnotesize
\begin{tabular}{lccc}
\toprule
Pattern & $n$ & Surface LR & \driftnet{} \\
\midrule
full hijack & 450 & 0.902 & 0.998 \\
partial hijack & 208 & 0.111 & 0.986 \\
delayed execution & 117 & 0.171 & 0.932 \\
\midrule
all attacked & 775 & 0.579 & 0.985 \\
\bottomrule
\end{tabular}
\end{table}

\subsection{Per-Domain and Per-Family Slices}
\label{sec:results-slices}

\begin{figure*}[!t]
\centering
\includegraphics[width=\textwidth]{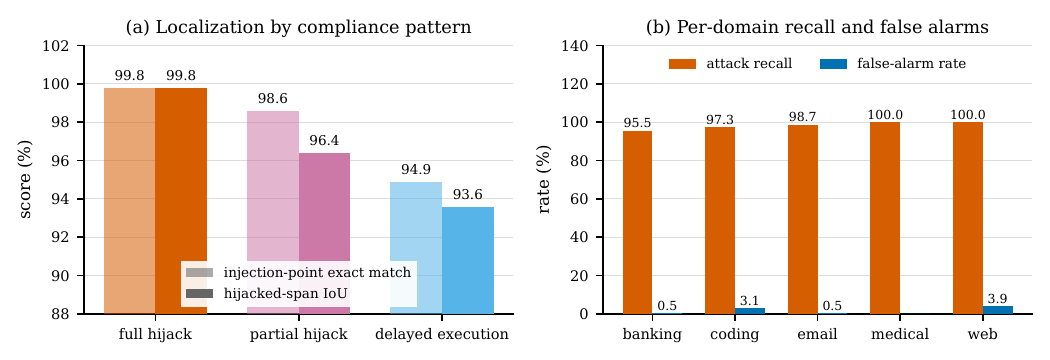}
\caption{(a) Localization by compliance pattern: injection-point exact match and hijacked-span IoU degrade mildly from full hijack to delayed execution. (b) Attack recall and false-alarm rate by domain; medical shows no false-alarm bar because its rate is exactly zero (0 of 209).}
\label{fig:locdomain}
\end{figure*}

Performance is high across all five of the benchmark's domains (Table~\ref{tab:domain}, Figure~\ref{fig:locdomain}). The medical domain is solved perfectly on this test part; banking has the lowest attack recall (0.955), contributing 5 of the 12 false negatives; web shows the highest false-alarm rate (0.039). Across the six attack-goal families, recall spans 0.961 (multi-step spreading) to 1.000 (branch divergence), so no attack goal is systematically missed; we report these figures descriptively, per the scope note of Section~\ref{sec:problem}.

\begin{table}[!t]
\caption{Held-out test by domain: trajectory counts, attack recall, false-alarm rate over non-attacked trajectories, and F1. Lower block: attack recall by attack-goal family; three trajectories tagged with the variant spelling \lbl{data\_theft} are counted under data stealing.}
\label{tab:domain}
\centering
\footnotesize
\begin{tabular}{lcccc}
\toprule
Domain & $n$ & Recall & False alarms & F1 \\
\midrule
banking & 317 & 0.955 & 0.005 & 0.973 \\
coding & 382 & 0.973 & 0.031 & 0.971 \\
email & 343 & 0.987 & 0.005 & 0.990 \\
medical & 376 & 1.000 & 0.000 & 1.000 \\
web & 304 & 1.000 & 0.039 & 0.981 \\
\midrule
\midrule
Attack-goal family & \multicolumn{2}{c}{$n$} & \multicolumn{2}{c}{Recall} \\
\midrule
branch divergence & \multicolumn{2}{c}{107} & \multicolumn{2}{c}{1.000} \\
reasoning corruption & \multicolumn{2}{c}{129} & \multicolumn{2}{c}{0.992} \\
parameter manipulation & \multicolumn{2}{c}{151} & \multicolumn{2}{c}{0.987} \\
data stealing & \multicolumn{2}{c}{134} & \multicolumn{2}{c}{0.985} \\
direct harm & \multicolumn{2}{c}{151} & \multicolumn{2}{c}{0.980} \\
multi-step spreading & \multicolumn{2}{c}{103} & \multicolumn{2}{c}{0.961} \\
\bottomrule
\end{tabular}
\end{table}

\section{Error Analysis}
\label{sec:errors}

The single test evaluation leaves 26 errors in 1{,}722 trajectories: 12 false negatives and 14 false positives. Both sets are small enough to read exhaustively, and both carry a lesson.

\subsection{The Twelve Misses}

Table~\ref{tab:fn} lists every false negative. The misses concentrate where detection is intrinsically hardest: 8 of 12 are delayed executions and 3 are partial hijacks, so 11 of 12 come from the two patterns in which one or two corrupted steps hide inside an otherwise normal trajectory; only one full hijack is missed. By domain they split across banking (5), coding (5), and email (2); medical and web attacks are never missed.

\begin{table}[!t]
\caption{All twelve false negatives: compliance pattern, attack-goal family, and the model's predicted attack probability.}
\label{tab:fn}
\centering
\scriptsize
\setlength{\tabcolsep}{2.5pt}
\begin{tabular}{lllc}
\toprule
Trajectory & Pattern & Family & $\hat{p}$ \\
\midrule
\lbl{banking\_attacked\_delayed\_0128} & delayed & data stealing & 0.430 \\
\lbl{banking\_attacked\_delayed\_0063} & delayed & direct harm & 0.110 \\
\lbl{coding\_attacked\_delayed\_0080} & delayed & multi-step spreading & 0.055 \\
\lbl{email\_attacked\_delayed\_0131} & delayed & data stealing & 0.007 \\
\lbl{email\_attacked\_delayed\_0001} & delayed & parameter manipulation & 0.004 \\
\lbl{banking\_attacked\_delayed\_0171} & delayed & direct harm & 0.002 \\
\lbl{banking\_attacked\_partial\_0283} & partial & reasoning corruption & 0.001 \\
\lbl{banking\_attacked\_delayed\_0072} & delayed & direct harm & 0.001 \\
\lbl{coding\_attacked\_partial\_0018} & partial & multi-step spreading & 0.001 \\
\lbl{coding\_attacked\_full\_0176} & full & multi-step spreading & 0.000 \\
\lbl{coding\_attacked\_partial\_0196} & partial & multi-step spreading & 0.000 \\
\lbl{coding\_attacked\_delayed\_0052} & delayed & parameter manipulation & 0.000 \\
\bottomrule
\end{tabular}
\end{table}

Two properties of the misses stand out. First, they are confident, not marginal: 11 of the 12 receive predicted attack probability below 0.12, and only one (at 0.43) sits near the threshold, so no threshold adjustment recovers them. Second, reading the twelve trajectories shows why. In 11 of the 12, the hijacked actions are internal operations with no world-feature signature: a card frozen, an email moved to a folder, a test run, a beneficiary list fetched, a command like \lbl{echo \$PATH} executed. More telling, in 9 of the 12 the observation labeled \lbl{injection\_point} contains no legible injected instruction at all; the observation reads as a routine tool result, and the attack's only textual trace, when there is one, appears in a later step. Those 9 misses are therefore at least partly attributable to residual generation noise in the corpus, consistent with the dataset paper's estimate of about 0.4\% label error~\cite{agentdrift2026dataset}, rather than to a detector failure on legible evidence: where the trajectory text carries no attack, there is nothing for any text-reading detector to find.

\subsection{The Fourteen False Alarms}

\begin{figure}[!t]
\centering
\includegraphics[width=\columnwidth]{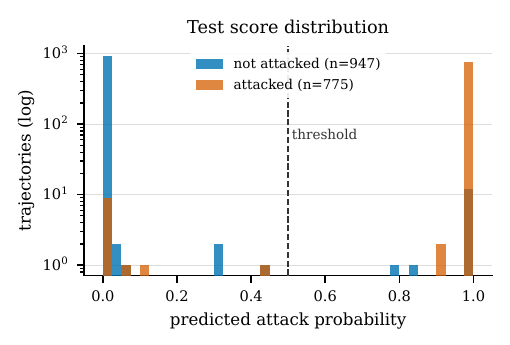}
\caption{Predicted attack probability on the held-out test part, log scale. Errors sit at the confident extremes, not near the threshold.}
\label{fig:scores}
\end{figure}

The false positives split into 8 benign trajectories and 6 hard negatives, concentrated in coding (6) and web (6). They are as confident as the misses: 12 of 14 receive probability above 0.99. The web cases are consistent with the world-feature channel's one blind spot: legitimate browsing also touches URLs outside the user's known domains, so a benign trajectory such as one that opens an external site and extracts its links presents the same external-URL signature as an attack. The coding cases split evenly between hard negatives and benign trajectories whose routine operations (security patches applied benignly, cleanup commands, log inspection) read like post-compromise activity. Figure~\ref{fig:scores} shows the full score distribution: the two classes separate almost completely, with a thin bridge of errors at both extremes rather than a mass near the threshold.

\section{Discussion}
\label{sec:discussion}

\subsection{Support for the Drift Hypothesis}

The results support the drift hypothesis directly, and the evidence is in the pattern-level structure rather than the headline. The detector's advantage over the surface baseline concentrates precisely where sequence reading is required: partial hijacks (0.111 to 0.986) and delayed executions (0.171 to 0.932); attacks with no lexical fingerprint; resisted attacks whose poison text is present but inert; and hard negatives whose surface resembles an attack while every action serves the user's task. A model that had merely learned sharper keyword statistics would show the opposite signature, gaining recall at the cost of false alarms on exactly those categories; instead every false-alarm rate falls while recall rises from 0.579 to 0.985.

\subsection{Localization Turns Flags into Actions}

Localization is what turns detection into something an operator can act on, and the strict metrics show it is dependable: the injection-point set is recovered exactly in 98.7\% of attacked trajectories and the hijacked span at IoU 0.979. In operational terms, a flagged trajectory arrives with the step to roll back to, the actions to audit, and the content source to distrust. This output is strictly richer than what contemporaneous step-level systems produce: a single onset or error index~\cite{felicia2026stepshield,trajad2026} conflates the poisoned read with the corrupted writes, and binary per-action judgments~\cite{zheng2026stepguard} cannot separate the two either, nor represent an agent that recovered after partial compliance or resisted the attack entirely. The perfect separation of \lbl{failed\_injection} steps, matching the zero trajectory-level flags on failed attacks, indicates the model has internalized the distinction the benchmark enforces: injected text is not compromise; compliance is.

\subsection{Why a Small Frozen-Encoder Model Suffices}

Everything expensive in \driftnet{} is frozen. The sentence encoder runs once per corpus, its embeddings are cached, and the trained component is under two million parameters; the sweep shows every drawn depth and width landing inside a 0.011 band of F1, so the capacity bottleneck on this benchmark is not the sequence model. This budget contrasts with the current trend toward LLM-scale guards: StepGuard fine-tunes a 4B-parameter model with reinforcement learning to emit binary per-action judgments~\cite{zheng2026stepguard}, and trajectory-guard benchmarks show frontier LLM judges reaching only 76.7\% F1 on trace-level safety~\cite{li2026atbench}. On a benchmark with dense supervision, a supervised sequence model three orders of magnitude smaller solves a strictly finer-grained task. The comparison is not like for like across datasets, but it does show that step-level injection triage does not intrinsically require an LLM-scale judge.

\subsection{A Calibration Caution for Deployment}

The error analysis carries one practical warning. The model's errors are confident: misses receive probability near 0, false alarms near 1, and the training dynamics show scores hardening after accuracy has plateaued. A deployment that consumes the raw probability (for alert ranking, or thresholding at an operating point other than the default) should calibrate on a held-out stream rather than trusting the training-time confidences, and should treat the score as a decision, not a degree of belief. The flip side is benign: because errors are not marginal, the operating point is insensitive to the threshold over a wide range.

\section{Limitations}
\label{sec:limitations}

\textbf{Synthetic, single-generator data.} All trajectories, attacked and benign alike, come from one generator model under one protocol~\cite{agentdrift2026dataset}. The task-disjoint split removes task memorization, but it cannot rule out reliance on generator-specific style, and this paper establishes nothing about transfer to other generators or to real agent traffic. The results are specific to this generator's distribution.

\textbf{World-identity regularity in the corpus.} The dataset paper measures a world-identity artifact in the benchmark: trajectories from the same generated world correlate with the same category, strongly enough that a lookup from world identity to majority training label reaches 86.1\% binary accuracy on its stratified test split~\cite{agentdrift2026dataset}; repeating that measurement on the task-disjoint split used here gives 86.9\% against a 55.0\% majority class. \driftnet{}'s four world features are immune by construction, encoding only internal-versus-external facts and never which world a trajectory belongs to, but the frozen text embeddings read the raw step text, which contains world-identifying strings such as names, companies, and addresses, so part of the trajectory-level headline could in principle ride on that regularity rather than on behavior. The step-level results cannot be explained this way, because localization is a within-trajectory prediction that world identity does not determine, and the near-zero flag rates on failed attacks and hard negatives are better than the lookup's accuracy could deliver; the honest statement is nevertheless that binary test numbers on this corpus should be read alongside that measured artifact, and the anonymized and world-held-out evaluations the dataset paper recommends are the right follow-up.

\textbf{Short trajectories.} The corpus spans 3 to 11 steps with mean 5.67. Behavior on much longer agent runs, where drift may unfold gradually and positional priors weaken, is untested.

\textbf{World-feature dependence.} Four input features presuppose a contact world that decides whether a recipient or URL is external. In deployments without such grounding the model would run on text semantics alone, and we have not measured that degradation; the web-domain false alarms in Section~\ref{sec:errors} already show the world-feature channel's blind spot on legitimately external content.

\textbf{Frozen encoder.} The detector reads fixed \lbl{all-mpnet-base-v2} embeddings. A jointly fine-tuned encoder might capture injection-relevant nuance the frozen one misses; we did not explore this, and doing so would sacrifice the precomputed-cache determinism that makes the runs cheap and their inputs exactly reproducible.

\textbf{No adaptive attacker.} The evaluation covers the benchmark's fixed attack distribution. An adversary aware of the detector could attempt injections whose induced behavior mimics benign drift or whose actions avoid the world-feature signature, as the low-signal miss cases already suggest; robustness to detector-aware attacks is unmeasured.

\section{Future Work}
\label{sec:future}

\textbf{Leakage-controlled evaluation.} Re-evaluating under the dataset paper's recommended protocols, anonymized surface forms and world-held-out splits, would bound how much of the trajectory-level headline survives with the world-identity regularity removed.

\textbf{Joint encoder fine-tuning.} Fine-tuning the step encoder with the trajectory model trades cached-embedding determinism for representation quality, and is the most direct route to gains if the frozen embeddings are the bottleneck.

\textbf{Multi-generator data and real traffic.} Broadening the corpus with trajectories from multiple generators would extend the task-disjoint discipline to a generator-disjoint one, and the decisive test remains evaluation on logged real agent traffic, alongside an LLM-judge comparison to quantify what a small supervised model gives up, or does not, against zero-shot flexibility.

\textbf{Streaming operation.} \driftnet{} scores a completed log, but nothing in the architecture requires completion: scoring growing prefixes would turn the same model into an online monitor, with the step head flagging the injection point as soon as the first hijacked action follows it.

\textbf{Longer horizons.} On longer agent runs the injection-to-execution gap can widen well beyond this corpus's spans; whether bidirectional attention over hundreds of steps sustains the delayed-execution results is open.

\section{Conclusion}
\label{sec:conclusion}

We presented \driftnet{}, a detector built on the premise that indirect prompt injection is legible in an agent's own behavior: read the trajectory in order, against the task and the world it acts in, and a successful attack appears as drift. The architecture follows the premise with deliberate economy. A frozen sentence encoder and four identity-free world features turn each logged step into a vector; a padding-masked Transformer trunk of at most three layers, under two million parameters in total, conditions every step on the whole sequence; and two heads trained against a single class-weighted objective read the shared states, one deciding whether the trajectory is compromised, one labeling every step as benign, injection point, hijacked, or failed injection. That joint output, which to our knowledge no prior detector produces, converts a detection into a triage report: the observation to distrust, the span to roll back, and the injections the agent already resisted on its own.

The evidence was gathered under a protocol designed to be hard to fool. On the task-disjoint split, with all 20 sweep configurations converging into a 0.011 band of validation F1 and the held-out test part evaluated exactly once, \driftnet{} reaches trajectory-level F1 of 0.983, recovers the exact injection-point set in 98.7\% of attacked trajectories and the hijacked span at IoU 0.979, flags zero of 218 resisted attacks, and flags 2.9\% of hard negatives. The comparison that carries the argument is pattern-level: a surface baseline retrained on the identical split catches 90.2\% of full hijacks but 11.1\% of partial hijacks and 17.1\% of delayed executions, while \driftnet{} reaches 99.8\%, 98.6\%, and 93.2\% with every false-alarm rate lower at once. Gains concentrated exactly where one or two corrupted actions hide inside an otherwise normal trajectory are gains from reading the sequence, not from sharper surface statistics. The exhaustive error reading adds a finding about the corpus itself: most residual misses are trajectories whose labeled injection observation contains no legible instruction, so they bound label noise rather than detector capability.

The claim stays inside the evidence. The corpus is synthetic and single-generator, its measured world-identity regularity is reported next to the headline numbers, and nothing here speaks to real agent traffic or detector-aware attackers. What the results do establish is that dense step-level supervision changes what a defense can be: not an LLM-scale judge returning a verdict, but a small, cheap, model-agnostic sequence model that says precisely where a trajectory went wrong, and that holds its accuracy exactly on the stealthy patterns where surface signals fail. Anonymized and world-held-out re-evaluation, generator-disjoint corpora, and streaming operation over growing prefixes are the natural next steps.

\balance
\bibliographystyle{IEEEtran}
\bibliography{references}

\end{document}